# Multi-Stakeholder Alignment in LLM-Powered Collaborative AI Systems: A Multi-Agent Framework for Intelligent Tutoring

Alexandre P Uchoa[0000-0001-7971-496X], Carlo E T Oliveira[0000-0003-4387-9732], Claudia L R Motta[0000-0002-4069-1462], Daniel Schneider[0000-0003-2987-4732]

Universidade Federal do Rio de Janeiro, NCE, Cidade Universitária, Rio de Janeiro RJ 21941-590, Brazil
secretaria@ppgi.ufrj.br

**Abstract.** The integration of Large Language Models into Intelligent Tutoring Systems presents significant challenges in aligning with diverse and often conflicting values from students, parents, teachers, and institutions. Existing architectures lack formal mechanisms for negotiating these multi-stakeholder tensions, creating risks in accountability and bias. This paper introduces the Advisory Governance Layer (AGL), a non-intrusive, multi-agent framework designed to enable distributed stakeholder participation in AI governance. The AGL employs specialized agents representing stakeholder groups to evaluate pedagogical actions against their specific policies in a privacy-preserving manner, anticipating future advances in personal assistant technology that will enhance stakeholder value expression. Through a novel policy taxonomy and conflict-resolution protocols, the framework provides structured, auditable governance advice to the ITS without altering its core pedagogical decision-making. This work contributes a reference architecture and technical specifications for aligning educational AI with multi-stakeholder values, bridging the gap between high-level ethical principles and practical implementation.

## 1 Introduction

Large Language Models (LLMs) will transform Intelligent Tutoring Systems (ITS) by enabling rich, context-aware dialogues with unprecedented pedagogical effectiveness. Yet, learning is not an isolated student-system interaction. It is embedded in a socio-educational context of diverse stakeholders—students, parents, teachers, and regulators— each with distinct values, preferences, and constraints that can fundamentally conflict and create a critical need for governance. This need for governance becomes even more critical as ITSs increasingly leverage AI to create educational content and decide students' progression autonomously, reducing the direct oversight of human educators and creating new risks of not properly scrutinized material. Governance is seldom a first-class concern in ITS design; policies are often hardcoded or omitted entirely. While bodies like UNESCO and the OECD provide ethical principles, no stand-ard architectural pattern exists to operationalize them. The challenge is to align educational AI with stakeholder values without sacrificing pedagogy or creating integration barriers.

LLMs and multi-agent systems offer novel solutions. Stakeholders can express policies in natural language, democratizing participation and soon may be represented by intelligent agents themselves. The architecture allows for flexibility via prompt engineering instead of code changes. A unified framework can handle policy interpretation, compliance checks, negotiation, and generate tailored explanations for diverse audiences, reducing system complexity. Ultimately, this approach positions the AGL not as a replacement for human judgment, but as a tool to amplify and scaffold the collective intelligence of those directly involved in the educational journey.

Despite these advances and the emergence of agent communication standards, no existing framework adequately addresses multi-stakeholder governance in educational AI. Current approaches lack mechanisms for explicit stakeholder representation, privacy-preserving policy evaluation, and transparent negotiation of conflicting values and preferences. Furthermore, they lack mechanisms to ensure the stakeholder inputs themselves are ethical and that the negotiation process is fair. This gap between high-level ethical principles and on-the-ground technical implementation is what motivates our research.

### 1.1   Research Questions

This paper is guided by two fundamental research questions aimed at bridging the gap between high-level ethical principles and practical implementation:

**RQ1:** How can a non-intrusive governance layer be architected to oversee Intelligent Tutoring Systems without altering its core pedagogical logic?

This question investigates the core design patterns—including agent roles, privacy-preserving data flows, and integration points—for an advisory system that augments an ITS without altering its core pedagogical logic, ensuring broad compatibility.

**RQ2:** How can diverse, natural-language stakeholder policies be operationalized for AI governance?

This question tackles the challenge of translating prose-based rules from parents, teachers, and institutions into a structured taxonomy and vocabulary that an LLM-powered system can reliably interpret and enforce. The remainder of this paper is structured as follows. Section 2 reviews the background literature on LLMs, multi-agent systems, and the challenges of AI governance in education. Section 3 details the proposed AGL framework, including its reference architecture, agent roles, governance lifecycle, and policy taxonomy. Section 4 presents a forward-looking research agenda, while section 5 addresses the research questions. Finally, Section 6 discusses the work's limitations and concludes with its future research.

## 2   Background

Large Language Models (LLMs) are transformative technologies for personalized learning due to their unique capabilities [44]. Built on massive datasets, they possess broad, multidisciplinary knowledge and can generate coherent, audience-calibrated text through contextual understanding, translation, and sentiment analysis [7, 41, 70]. Beyond knowledge retrieval, LLMs demonstrate complex reasoning skills, including problem decomposition, creative solution generation, and planning. Crucially, their ability to adapt dynamically via in-context learning (ICL) allows them to acquire new knowledge and perform novel tasks without retraining, offering unprecedented flexibility [2, 19, 25, 47–49, 52, 84, 89, 92].

LLMs have enabled researchers to conceive and design LLM-powered agent frameworks that exhibit reasoning and planning skills comparable to symbolic agents, but without requiring handcrafted rules or reinforcement learning that make generalization challenging [54]. Because they can operate over arbitrary text, LLM-based agents are more flexible and adaptable than logic-based systems [14, 30, 86, 93]. LLM-based agents also have a larger field of perception and action what allow them to employ multimodal perception and tool utilization to acquire new knowledge and resources [59, 69, 75] either by learning from feedback or by performing new actions [68]. LLM-based agents can consistently embody different roles and perspectives when given appropriate prompts, interact seamlessly with each other in natural language, whether collaborating or competing , and coordinate on complex tasks [34, 51]. However, this adaptability also means that LLM-based multi-agent systems can exhibit concerning emergent behaviors, reflecting the unpredictability of the underlying models themselves [97, 98].

A pressing challenge is the capacity for LLMs to be 'deceptively aligned' [31, 66, 74], a trait that can persist through safety training [36, 64] and manifest as strategic 'Machiavellian' behaviors [65]. Beyond behavioral deception, these models are also vulnerable to technical exploits like backdoor attacks [46] and struggle to operate reliably in deceptive environments where they must counter misinformation [87]. These problems are compounded by the unpredictability of scaling: as models grow, they not only exhibit emergent abilities [53] but also introduce emergent risks [90], including the development of long-term planning capabilities that could circumvent human control [15].

Defining and achieving alignment remains a central challenge, as there is no universal consensus on its meaning. There is also the fundamental problem of "whose values should AI align with?" The challenge of value alignment is acutely critical in education, where AI systems must navigate competing curricula, diverse cultural values, and different philosophical perspectives on learning [56, 63]. Educational settings inherently involve multiple stakeholders—from students and parents to teachers and administrators—each with distinct definitions of value and competing objectives [4]. Consequently, designing effective adaptive learning is a balancing act between student, domain, and pedagogical models, all of which encode different stakeholder aims and trade-offs [67]. Personalization cannot be divorced from this broader ecosystem of curricula and cultural norms if it is to serve all parties effectively [44]. These overlapping concerns generate fundamental tensions, such as adaptability to student interests versus adherence to a curriculum [72], engaging content versus scaffolded progression [50], and accommodating family values while maintaining institutional integrity [4].

Reflecting on these opportunities, numerous intelligent learning systems are incorporating LLMs, but primarily for pedagogical functions like natural-language interaction, adaptive feedback, and student engagement. While these systems demonstrate the utility of LLMs for instruction and communication, their architectural focus remains on the student-tutor dyad, largely omitting the multi-stakeholder aspect of education [28] and introducing significant governance challenges, including opaque decision flows, emergent behaviors, and the risk of bias or privacy leaks.

To guide the responsible use of AI in education, several international bodies have established ethical frameworks. Organizations like UNESCO, the OECD, IEEE, and the European Commission converge on a set of core principles for trustworthy AI [11, 23, 32, 33, 61, 83]. These principles consistently emphasize human-centric design and agency, ensuring AI empowers rather than displaces users. They call for robust safeguards against bias to promote equity and inclusion, and mandate transparency and explainability so that system logic is auditable. Furthermore, they require technical robustness and safety, rigorous data privacy protection, and clear lines of accountability to maintain trust and address systemic risks.

While these international frameworks articulate critical norms, a significant gap persists between principles and practice. Reviews of existing AI ethics tools reveal an overreliance on post-hoc explanation and a lack of usable mechanisms for real-time governance [24, 57]. Specifically, there are no concrete architectural patterns for enforcing these norms within complex, multi-agent educational systems.

Addressing these risks and governance challenges requires treating the multi-stakeholder educational environment as a problem of Computer-Supported Cooperative Work (CSCW). Foundational CSCW and human-AI teaming research establish that effective group coordination depends on shared communication channels, transparent negotiation processes, and robust audit trails [5, 17, 38, 39, 88].

This evolution from classic ITS meta-architecture to multi-agent frameworks, combined with proven governance patterns from CSCW and distributed systems, motivates our design. We argue for horizontal governance overlays rather than monolithic policy engines to better accommodate evolving stakeholder norms and reduce integration friction. Therefore, we propose the Advisory Governance Layer (AGL): a non-intrusive framework that combines conflicts adjudicating [22], the principles of a "watchdog interceptor" [42, 81] and "runtime monitoring" [20, 85, 91] to sit atop any existing ITS, aligning with recent calls for ethical, governable AI in education.

## 3    The AGL Reference Architecture

This section details the Advisory Governance Layer (AGL), our reference model for non-intrusive governance. The AGL implements a layered architecture of stakeholder representation, decision mediation, and oversight agents. Its core principle is a strict separation between governance advisement and pedagogical logic, which ensures alignment with stakeholder values and preferences without compromising educational effectiveness. This separation is a key risk mitigation strategy. The AGL is intentionally advisory rather than prescriptive, preserving the pedagogical autonomy of the core ITS—a crucial requirement for adoption by educators. So, even if a governance agent hallucinates or makes an erroneous assessment, its output is an advisory input to the ITS, not an executive command, preserving a critical layer of safety.

We opted for a distributed, multi-agent architecture over a single monolithic governance engine to ensure stakeholder privacy and policy autonomy, which we be-lieve is critical for families and competing institutions (see Figure 1). This non-intrusive, federated approach was chosen as a pragmatic alternative to more rigid, centralized control systems that would face higher barriers to real-world implementation. And, although its innovativeness, this design maintains compatibility with both legacy FIPA-based systems and modern LLM orchestration frameworks.

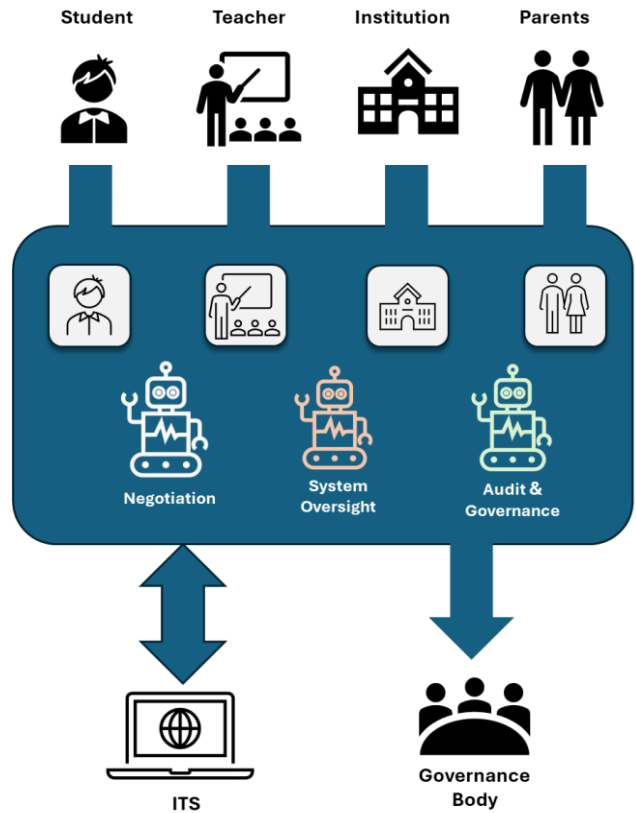

**Fig. 1.** High-level illustration of the AGL with its main components and its connection to the ITS and real counterparts.

### 3.1 Agent Roles and Responsibilities

The AGL's architecture distributes governance across four distinct agent types, separating policy evaluation from coordination and oversight.

**"Stakeholder Agents (SH)** represent the interests of single parties (e.g., Student, Parent, Institution, Regulator). Within the AGL architecture, they serve as the interface for stakeholder values and preferences. Crucially, these agents can either be internal components representing stakeholders or, more broadly, act as standardized proxies for external, third-party agents that stakeholders might employ in a future digital ecosystem. In either case, the AGL's role is to provide the interoperability layer for this communication.

Each SH agent is considered an autonomous policy engine that evaluates proposed ITS content options against its private repository of rules. Its primary function is to generate a structured vote (e.g., approve, reject) with justifications. This distributed design is crucial for privacy, allowing, for instance, a Parent agent to enforce family values without exposing them to the wider system.

**Multi-Stakeholder Negotiation Agent (MSN)** acts as the central coordinator. It does not access stakeholder policies directly; instead, it aggregates the private votes from all SH agents. Its core function is to orchestrate the governance process by identifying consensus and resolving conflicts using a configurable negotiation strategy. This ensures that core pedagogical values or safety constraints from one stakeholder can appropriately override the transient preferences of another. This allows an institution to implement a protocol that fits its specific governance needs, whether it be a strict hierarchical model, a weighted voting system, or a consensus-based approach. The MSN then synthesizes these inputs into a single, unified governance recommendation for the ITS. It is also responsible for generating a justification for its recommendation, explaining how stakeholder votes were reconciled based on the active negotiation protocol.

**Audit and Governance Agent (AG)** serves as the system's compliance officer and record-keeper. During the governance lifecycle, it passively records an immutable audit trail of all events, including links to the ITS activities, votes, justifications and negotiation outcomes. Crucially, its active analysis occurs post-hoc, outside the critical path of the decision-making loop, thus avoiding any performance bottleneck. After a decision is finalized, the AG inspects the transactional record for violations of system-wide principles that transcend any single stakeholder, such as unethical, biased, or malicious policies (e.g., a discriminatory preference submitted by a stakeholder), flagging them as part of the audit trail and generating tailored, multi-audience explanations of governance decisions for different stakeholders for transparency. This can then be also flagged for human review or trigger automated cross-validation requests to other agents.

**System Oversight Agent (SO)** operates at a meta-level, monitoring the long-term health and effectiveness of the governance system itself. Unlike the AG, which focuses on individual decisions, the SO analyzes aggregate data over time to detect systemic patterns, such as policy drift, stakeholder alert fatigue, or emergent agent collusion. While the AG Agent is a transactional auditor, the SO Agent is a systemic analyst. This also includes identifying long-term patterns of manipulative negotiation tactics or consistently biased policy submissions from any single stakeholder agent. This longitudinal oversight, with longer, historical contexts is designed to catch subtle, systemic biases or deceptive patterns that might be invisible in any single transaction, ensuring the system remains aligned with community values over the long term. When such issues are found, it generates reports for human review, preserving human authority over the evolution of the governance framework.

### 3.2   The AGL Governance Lifecycle

The AGL operates as a continuous, four-phase governance loop that provides post hoc advisory oversight without disrupting the ITS's pedagogical flow. The entire sequence is designed for real-time educational responsiveness, balancing thoroughness with low latency.

**1. Distributed Evaluation**: The cycle begins when the ITS Decision Engine generates a set of candidate actions (e.g., potential lessons or exercises) that is broadcasted to all active Stakeholder (SH) agents. Each SH agent then independently and privately evaluates these candidates against its local policies. It returns a structured vote (e.g., approve, reject, conditional) and a justification, but is not required to expose the underlying policy content. This federated evaluation respects stakeholder privacy boundaries, particularly for sensitive family values.

**2. Coordinated Recommendation**: The Multi-Stakeholder Negotiation (MSN) agent aggregates these distributed votes and justifications. If votes conflict, it applies to the institutionally configured resolution protocol. For example, in a deployment using a hierarchical strategy, a 'reject' vote from a *Regulator* would automatically supersede a 'conditional approval' from a *Teacher*. The MSN synthesizes the outcome into a single, unified governance assessment—including aggregated risk scores and highlighted concerns—and sends it back to the ITS as the advisory input.

**3. Autonomous Decision and Audit**: The ITS Decision Engine, retaining full pedagogical autonomy, makes the final selection, considering or not the advisory input provided by the AGL's assessment. Once the decision is made, the Audit and Governance (AG) agent captures the complete context—from candidate actions and stakeholder votes (including the MSN's justification for its recommendation) to the final outcome —in a secure, distributed audit trail. The AG then generates tailored explanations for different stakeholders, while the System Oversight (SO) agent begins analyzing the decision as part of a long-term data stream to detect systemic patterns.

**System Resilience**: The framework is designed for robustness. In cases of component failure, it degrades gracefully (e.g., proceeding with partial governance input) rather than blocking the ITS. For time-critical safety issues, emergency bypass protocols allow the ITS to act immediately while logging the event for post-hoc review, ensuring educational continuity is always prioritized."

**Table 1.** The proposed Policy Taxonomy and its four types.

| Type | Examples | LLM Processing Strategy |
|---|---|---|
| Hard Constraints | "No student data may be shared with third parties without explicit consent" (privacy regulation);<br>"All mathematics content must align with Common Core State Standards" (institutional policy);<br>"Students under 13 cannot access social media features" (safety constraint). | Few-shot prompting [8, 21, 78] extracts "must," "never," "required," and "prohibited" statements, converting them into Boolean predicates through semantic parsing and entity recognition [18, 94]. The Compliance Agent processes these as binary pass/fail evaluations. |
| Soft Preferences | "Prefer project-based learning activities when possible" (pedagogical preference);<br>"Minimize screen time before 6 PM" (parental guideline);<br>"Encourage collaborative rather than individual assignments" (teacher methodology). | Sentiment analysis and preference extraction [9, 41] identify stakeholder priorities, translating them into weighted scoring functions with configurable thresholds. Violations generate informational alerts but do not prevent action selection. |

| Temporal Rules | "Complete algebra fundamentals before introducing calculus concepts" (prerequisite sequencing); "No homework assignments after 8 PM on weekdays" (temporal constraint); "Provide reading support for students below grade level before advancing" (developmental timing). | Calendar integration combined with dependency parsing identifies time-based conditions and sequential requirements. The system evaluates candidate actions against temporal contexts and learning progression models [95]. |
| --- | --- | --- |
| Hierarchical Rules | "Institutional safety policies override parental content preferences" (authority precedence); "Federal accessibility requirements take precedence over efficiency optimizations" (regulatory hierarchy); "Teacher pedagogical decisions supersede student convenience preferences" (educational authority). | Ontology-based reasoning determines stakeholder authority levels and establishes conflict resolution precedence using predefined hierarchy mappings and contextual authority assessment [12, 58, 73]. |

### 3.3 Policy Taxonomy and Classification

To operationalize stakeholder values, the AGL requires a systematic method for interpreting diverse, natural-language policies. Drawing on analyses of AI governance in education [3, 10, 13, 16, 23, 26, 27, 40, 61, 62, 83], we developed a policy taxonomy that classifies rules into four principal categories (see Table 1). While drawing on existing classification schemes in AI governance, our taxonomy's contribution is its specific adaptation and formalization for the unique, multi-stakeholder conflicts inherent to educational settings. This framework is designed to accommodate everything from rigid regulatory mandates to flexible individual pedagogical preferences, enabling reliable translation into machine-executable logic. The four categories are:

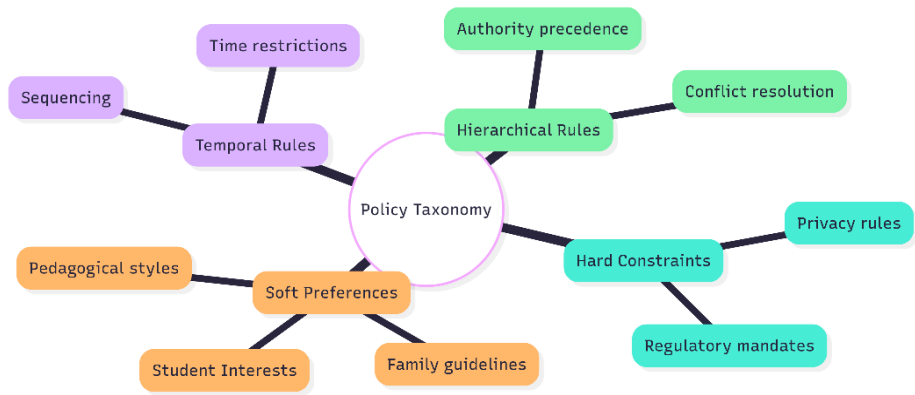

**Fig. 2.**   The adapted taxonomy for the multi-stakeholder conflicts inherent to educational settings, accommodating from rigid regulatory mandates to flexible individual pedagogical preferences.

**Hard Constraints**: These are non-negotiable, mandatory requirements, typically derived from regulations or core institutional policies (e.g., data privacy laws). They are processed as Boolean predicates where any violation results in a hard 'reject' vote.

**Soft Preferences**: These represent desirable but overridable guidelines, reflecting best pedagogical practices or individual stakeholder tastes and values (e.g., "prefer hands-on activities"). The system translates these into weighted scores that influence, but do not determine, the final governance recommendation.

**Temporal Rules**: These policies specify time-bound conditions, such as deadlines, content sequencing, or developmental prerequisites (e.g., "no homework over holidays"). They require temporal reasoning to evaluate whether an action is appropriate for a given context and time.

**Hierarchical Rules**: These are meta-rules that define precedence among other policies to resolve conflicts. They establish an authority structure (e.g., "regulatory constraints override institutional policies"), which is critical for the MSN agent's negotiation protocols.

The AGL uses LLM-driven techniques like few-shot prompting, semantic parsing, and entity recognition to extract and classify these rules from natural-language input [8, 9, 12, 21, 41, 58, 73, 78, 95].

**Policy Representation and Storage.** Each SH agent maintains its own policy repository using a standardized hybrid schema. This schema combines structured metadata (e.g., authority level, validity dates) with the original natural language expression of the policy. This dual representation supports both machine processing and human interpretability. A version control system tracks all policy modifications, ensuring a complete audit trail of how governance requirements evolve over time.

Policies are encoded using a hybrid representation that combines structured metadata with natural language content. The metadata layer captures policy attributes, including temporal validity, authority level, affected domains, and precedence relationships. The content layer preserves the original natural language expression while adding semantic annotations that facilitate processing. This dual representation supports both human interpretability and machine processing, essential for maintaining transparency in governance decisions.

Version control mechanisms track policy evolution over time, enabling retrospective analysis of how governance requirements change and supporting audit requirements. Each policy modification generates a new version with associated metadata explaining the change rationale and authority. This temporal dimension proves crucial for understanding governance decisions in context and evaluating their long-term effectiveness.

### 3.4 Privacy-Preserving Distributed Evaluation

The AGL's architecture operationalizes the policy taxonomy through a distributed evaluation model designed to protect stakeholder autonomy and privacy.

**Federated Evaluation and Privacy-Preserving Voting**. Rather than centralizing data, the AGL employs a federated approach where each Stakeholder (SH) agent evaluates ITS proposals locally against its private policies. Governance is achieved through the exchange of structured, privacy-preserving votes. These votes reveal only the outcome (e.g., approve/reject), a confidence score, and an opaque justification ID, preventing the inference of the underlying policy rules. Techniques like differential privacy can be applied to vote aggregation to further protect individual stakeholder positions from analysis.

**Policy-Agnostic Conflict Resolution**. When votes conflict, the Multi-Stakeholder Negotiation (MSN) agent resolves them without accessing the private policies themselves. It operates solely on vote metadata and the pre-defined hierarchical rules from the policy taxonomy. For instance, a 'reject' vote from a *Regulator* agent automatically takes precedence over a 'conditional approval' from a *Teacher* agent based on their relative authority, not the content of their policies. This ensures that negotiation respects authority structures while maintaining policy confidentiality.

**Distributed Audit and Performance**. To ensure accountability, the system generates a dual-layer audit trail. Each SH agent keeps a private, local log of its evaluations, while the Audit and Governance (AG) agent maintains a global ledger of the decision choreography (e.g., who voted, how conflicts were resolved), linked cryptographically to the local logs. This provides end-to-end verifiability without centralizing sensitive data. To meet real-time educational needs, the architecture uses performance optimizations such as caching recent evaluations, pre-computing likely decisions, and employing adaptive timeouts to balance evaluation thoroughness with system responsiveness.

### 3.5   Implementation Foundations and Integration

**Governance Hook Specifications**. To ensure broad compatibility, the AGL integrates with an ITS via a standardized event taxonomy called Governance Hooks. This approach provides a uniform governance fabric without altering the ITS's internal logic [35, 45, 79, 81, 82]. We define five hook categories:

1) **Decision Hooks** fire when the ITS proposes or selects actions (e.g., onCandidatesProposed);

2) **Policy Hooks** trigger rule changes (e.g., onPolicyUpdated);

3) **Interaction Hooks** capture human-in-the-loop events (e.g., onOverrideRequested);

4) **Exception Hooks** signal compliance failures (e.g., onCandidateRejected); and

5) **Periodic Hooks** support scheduled audits (e.g., onWeeklyDriftScan). This event-driven model allows the AGL to non-intrusively monitor any ITS (see Table 2).

**Integration Patterns and APIs**. Integration is operationalized through a minimal set of APIs that leverage established standards. The AGL requires endpoints for broadcasting candidate actions, delivering governance responses, and querying audit trails. All communications use standardized JSON schemas for interoperability, with policy representations informed by XACML principles and audit trails using the W3C PROV ontology [100, 102]. This allows the AGL to be implemented as a non-intrusive middleware layer. For legacy FIPA-ACL systems, it translates performatives into hook eventsFor modern LLM frameworks like LangChain, it wraps the decision-making chain to add pre- and post-governance checks.

**Table 2.** Hook categories and examples of events they link to.

| Hook Category | Definition | Example Event Types | Origin / Mapping to Established Concepts |
|---|---|---|---|
| Decision Hooks | Emitted when the ITS decision engine presents candidate actions or selects a final action. | onCandidatesProposed(), onDecisionChosen() | Aligns with XACML's Policy Decision Point (PDP) evaluations and FIPA-ACL "propose/inform" performatives. |
| Policy Hooks | Fired when stakeholders create, update, version, or retire governance policies and rules. | onPolicyUploaded(), onPolicyVersioned(), onPolicyExpired() | Mirrors XACML's Policy Administration Point (PAP) and Rei/KAoS policy-change notifications. |
| Interaction Hooks | Triggered by human-in-the-loop actions: overrides, acknowledgements, feedback submissions. | onOverrideRequested(), onAlertAcknowledged(), onFeedbackSubmitted() | Corresponds to GDSS override events and FIPA-ACL "request/confirm" performatives. |
| Exception Hooks | Indicate hard rejections or missed obligations (e.g., candidate rejections, unacknowledged alerts). | onCandidateRejected(), onOverrideDenied(), onAcknowledgementMissed() | Reflects XACML PDP "deny" responses and CEP "non-event" patterns. |
| Periodic Hooks | Scheduled or rolling-window audits and scans (e.g., fatigue checks, policy drift analyses). | onDailyAudit(), onWeeklyFatigueCheck(), onMonthlyDriftScan() | Drawn from CEP sliding/tumbling-window queries and runtime verification polling. |

These hooks provide comprehensive governance coverage while maintaining architectural flexibility. Legacy FIPA-ACL systems integrate through message interceptors that translate performatives into standardized events, while modern LLM frameworks leverage native extension mechanisms for event emission.

**Incremental Deployment.** Organizations can adopt the AGL incrementally. Phase one can implement basic audit logging. Phase two can introduce real-time policy evaluation for a subset of decisions. Phase three can then enable the full distributed governance framework. This phased approach reduces implementation barriers and allows organizations to realize value at each step.

### 3.6 A Running Vignette

The following scenario demonstrates how the AGL navigates the complex intersection of pedagogy, equity, and conflicting stakeholder values (see Figure 3).

**Scenario:** An ITS is tutoring Student X, a 9th-grader from a low-income household who struggles with abstract math but excels with hands-on learning. The ITS generates three candidate geometry lessons: (1) Abstract proofs, (2) Construction/trade-focused applications, and (3) Art-integrated design.

**1. Stakeholder Evaluation:** Each SH agent privately evaluates the candidates.

The *Student Agent* favors options 2 and 3 based on a preference for practical relevance (Soft Preference).

The *Teacher Agent* also favors option 2 for its alignment with the student's learning style but requires it to maintain academic rigor (Hard Constraint).

The *Parent Agent*, concerned about stereotypes, flags option 2 for potentially steering the student away from academic tracks based on socioeconomic background.

The *Regulator Agent* issues a 'reject' vote for option 2, citing a policy against perpetuating socioeconomic tracking (Hard Constraint).

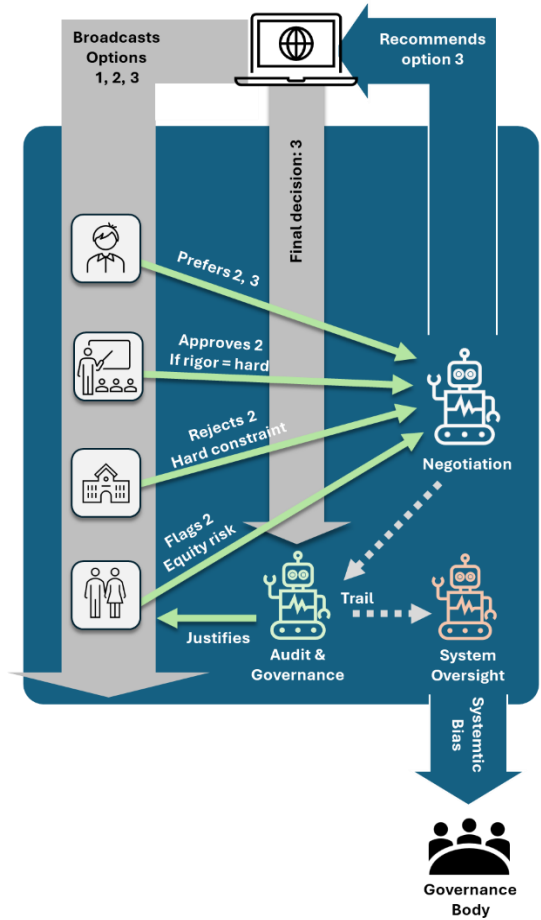

**Fig. 3.**   Illustration of the running vignette where AGL's core agents evaluate, explain decisions to stakeholders, record decision trail and provide long-term oversight over the behavior of all components, including stakeholder agents.

**2. Negotiation and Recommendation**: The MSN agent receives conflicting votes. In this scenario, the MSN is configured with a hierarchical negotiation protocol where regulatory rules have the highest precedence. Consequently, the *Regulator's* hard constraint automatically disqualifies option 2. The MSN then synthesizes a recommendation for option 3 (Art-integrated geometry), as it best balances pedagogical effectiveness (satisfying the *Student* and *Teacher*) with the critical equity concerns (addressing the *Parent* and *Regulator*). This unified governance assessment is then forwarded to the ITS.

**3. Decision, Audit, and Oversight**: The ITS, retaining full autonomy, selects the recommended art-integration approach. The AG agent immediately logs the entire decision trail and generates tailored explanations. The transparent audit record now includes the MSN's justification that regulatory rules on equity took precedence. For the *Parent*, it explains how the choice respected their equity concerns while maintaining academic rigor.

Weeks later, the SO agent's longitudinal analysis detects a subtle pattern: students from similar demographics are consistently steered away from abstract reasoning across the system. It flags this systemic bias for human review, demonstrating how the AGL provides oversight that transcends any single decision.

## 4   Open Research Challenges

The AGL provides a foundational architecture for multi-stakeholder governance, yet its implementation in eliciting and reconciling stakeholder values and preferences surfaces several critical research challenges that must be addressed to realize robust, practical systems.

**Dynamic Policy Extraction**. Converting prose-based institutional policies into executable rules as well as eliciting and formalizing dynamic stakeholder preferences is a primary challenge. While LLMs can decompose policy snippets [21, 37], extending these methods to handle large, nested legal and institutional documents reliably is an open problem. Future work must develop mixed-initiative systems where LLMs draft candidate rules, but human stakeholders can verify, refine, and formally approve them through intuitive interfaces that manage cognitive load.

**Collusion-Resilient Governance.** A novel risk in multi-agent systems is that multiple agents sharing the same underlying LLM could inadvertently or maliciously collude, for instance by homogenizing their outputs [6], or coordinating to bypass compliance checks. While some work explores coalition formation [71] and monitoring [29], runtime defenses against such coordinated manipulation are underdeveloped. A critical challenge is to design architectural patterns and anomaly-detection algorithms for the AG or SO agents to spot and mitigate these synchronized deviations in real time.

**Trust-Centered Alert Design & Stakeholder Communication.** For the AGL to be effective, its governance alerts must be understood, trusted, and actionable by non-technical stakeholders like parents and teachers. However, as prior work on trust [1] and cognitive load [55] suggests, poorly designed notifications can quickly lead to alert fatigue and mistrust. A key research agenda is to determine the optimal design for governance alerts, investigating how factors like linguistic style, empathetic framing, frequency, and information density impact stakeholder comprehension, trust, and willingness to intervene.

**Privacy-Preserving Governance Pipelines**. A governance layer that inspects raw learner data risks violating privacy regulations like GDPR. The challenge is to verify rule satisfaction without exposing personally identifiable information. While foundational techniques like differential privacy (DP) [43, 60], zero-knowledge proofs (ZKPs) [76, 101], and federated evaluation [77] exist, significant work is needed to adapt them for real-time educational governance [80, 96, 99]. Optimizing DP-noise calibration, reducing ZKP verification costs, and minimizing federated communication overhead are all critical research frontiers.

**Design and Evaluation of Negotiation Protocols.** The AGL's MSN agent is designed with a configurable negotiation module, but the optimal strategy for resolving stakeholder conflicts is an open question. Research is needed to design and empirically evaluate different conflict resolution protocols (e.g., hierarchical, consensus-based, weighted voting). Key questions include: Which protocols best foster stakeholder trust? How do different strategies impact educational equity? And can protocols be made provably fair or resistant to manipulation, not only from colluding agents but also from individual bad-faith actors submitting unethical policies?

## 5    Addressing the Research Questions

This paper was guided by two fundamental research questions. The AGL framework, as detailed in the preceding sections, provides concrete answers to both.

**RQ1: How can a non-intrusive governance layer be architected to oversee multi-agent Intelligent Tutoring Systems?**

This paper answers RQ1 by proposing the Advisory Governance Layer (AGL), an architecture founded on non-intrusive oversight. The key architectural patterns enabling this are: (1) Separation of Concerns, where the AGL operates as an advisory overlay, preserving the ITS's pedagogical autonomy; (2) Distributed Agent Roles (SH, MSN, AG, SO), which decompose governance into distinct, manageable functions; (3) A Privacy-Preserving Lifecycle, which incorporates stakeholder input without exposing sensitive policy content; and (4) A Standardized Integration Fabric via Governance Hooks, which allows the AGL to monitor any ITS without requiring brittle, deep integration.

**RQ2: How can diverse, natural-language stakeholder policies be operationalized for AI governance?**

Our answer to RQ2 is a systematic, two-part solution for translating abstract values into enforceable rules: (1) A Policy Taxonomy, which provides the essential semantic structure by classifying rules as Hard Constraints, Soft Preferences, Temporal, or Hierarchical, thereby allowing the system to interpret their intent; and (2) A Hybrid Representation Approach, which stores policies in a dual format combining structured metadata for automated processing with the original natural-language text for human-centric transparency and auditability. This strategy creates a reliable pipeline for operationalizing stakeholder values within the AGL.

## 6    Discussion and Conclusion

The integration of LLMs into education presents a dual challenge: harnessing their potential for personalization while governing the risks of multi-stakeholder value conflicts. This paper introduced the Advisory Governance Layer (AGL), a non-intrusive, multi-agent framework designed to address this challenge. By separating pedagogical decision-making from governance advisement, and by enabling distributed, privacy-preserving stakeholder participation, the AGL provides a concrete architectural pattern for bridging the gap between high-level ethical principles and their practical implementation in live educational systems. However, we acknowledge several limitations that point toward essential future work.

**Limitations.** We acknowledge that the primary limitation of this work is its theoretical nature. The AGL is presented as a conceptual architecture, and its core claims of effectiveness and non-intrusiveness have not yet been empirically validated through user studies, simulations, or prototype implementation. Furthermore, while the framework provides the structure for governance, its successful application depends on robust underlying knowledge infrastructure and careful consideration of cultural variations in learning values. Finally, the balance between transparency and cognitive overload for stakeholders remains a critical human-computer interaction challenge; excessive alerts or overly complex explanations, as well as automatic, not reported judgements and choices could undermine the very trust the system aims to build. Conversely, the AGL's transparency is designed to empower this essential human oversight. By providing clear audit trails and justifications, the framework aims to enable stakeholders to question, understand, and even counter-argue governance assessments, ensuring they remain active participants who can correct for AI fallibility. We also acknowledge the inherent difficulty in having an AI accurately interpret and enforce 'soft' semantic rules, a task that may require ongoing human-in-the-loop validation, as noted in our research agenda.

**Validation Plan.** The open research challenges identified in this paper (see 4. Open Research Challenges)—from dynamic policy extraction and collusion resistance to the design of fair negotiation protocols—form a clear agenda for advancing the field. The necessary next step is the empirical validation of this framework. We propose a research plan proceeding in stages:

*Simulation Studies*: Simulate the governance lifecycle under various conflict scenarios to analyze the performance and fairness of different negotiation protocols compared to LLM-based agentic preference elicitation, dialoguing and negotiating setups.

*Prototype Implementation*: Develop a prototype of the AGL agents and hooks to test the feasibility of the architecture with an existing open source ITS.

*Controlled User Studies*: Conduct studies with teachers, parents, and students to measure the impact of the AGL on trust, cognitive load, and perceived fairness (as outlined in our research agenda).

*Scalability Studies*. Analysis of system performance under realistic loads, including thousands of concurrent students and millions of daily decisions, to validate architectural assumptions.

*Integration Standards*. Development of standardized APIs and governance protocols that enable vendor-neutral implementations, possibly through partnerships with educational technology standards bodies.

As AI systems shape the educational experiences of millions of students, our governance choices today will determine whether these technologies amplify existing inequities or help create more just and effective educational systems. The AGL is not a panacea for AI fallibility but a framework for structured collaboration, designed to make the entire socio-technical system more intelligent and accountable, even when its individual components—human or AI—are imperfect. The time for principled, implementable governance frameworks is now.